\documentclass{PoS}

\title{Results from IceCube}

\ShortTitle{ICRC 2019 Results from IceCube}

\author{
The IceCube Collaboration\footnote{For collaboration list, see PoS(ICRC2019) 1177.}\\
{\itshape \href{http://icecube.wisc.edu/collaboration/authors/icrc19_icecube}{http://icecube.wisc.edu/collaboration/authors/icrc19\_icecube}}\\
E-mail: \email{drwilliams3@ua.edu}
}

\abstract{The IceCube Neutrino Observatory is the world's largest neutrino detector, instrumenting a cubic kilometer of ice at the geographic South Pole. The detector probes neutrino energies from GeV to PeV, and collects high statistics neutrino samples by virtue of its extremely large volume. IceCube was designed to detect high-energy neutrinos from potential cosmic ray acceleration sites such as active galactic nuclei, gamma ray bursts and supernova remnants. IceCube announced the detection of a diffuse flux of astrophysical neutrinos in 2013, including the highest energy neutrinos ever detected. IceCube has since developed a robust system of realtime alerts generated by astrophysical neutrino candidates that trigger rapid follow-up observations by other telescopes and detectors. In September 2017, IceCube observed a neutrino in coincidence with a flaring blazar, signaling the era of neutrino astronomy and widening the field of multi-messenger astronomy. The IceCube science program extends beyond multi-messenger astronomy to include the fundamental physics of neutrino oscillations and neutrino interactions with matter, cosmic ray physics, and searches for particles and forces beyond the Standard Model. This talk will cover the latest neutrino physics and astrophysics results from IceCube and future prospects for the detector. \\

\vspace{4mm}
{\bfseries Corresponding authors:}
\speaker{Dawn R. Williams}$^{1}$\\
{$^{1}$ \itshape University of Alabama}

}

\FullConference{36th International Cosmic Ray Conference -ICRC2019-\\
		July 24th - August 1st, 2019\\
		Madison, WI, U.S.A.}

\newcommand{\txs}{TXS~0506+056}
\newcommand{\numu}{$\nu_{\mu}$ }
\newcommand{\nue}{$\nu_{e}$ }
\newcommand{\nutau}{$\nu_{\tau}$ }
\newcommand{\mydeg}{$^{\circ}$ }

\begin{document}
\section{Introduction}
The IceCube Neutrino Observatory~\cite{Aartsen:2016nxy} is a
cubic-kilometer high-energy neutrino detector built at the geographic
South Pole near the Amundsen-Scott South Pole Station. The detector
consists of 86~cables called ``strings'', each instrumented with
60~Digital Optical Modules (DOMs) deployed between 1450~m and 2450~m
deep in the glacial ice. The DOM is a glass pressure vessel containing
a 10-inch photomultiplier tube (PMT)~\cite{Abbasi:2010vc} and
digitizing electronics~\cite{Abbasi:2008aa}, as well as 12
LED~flashers for calibration. The central, densely spaced ``DeepCore''
subarray~\cite{Collaboration:2011ym} is equipped with high quantum efficiency PMTs to lower the
neutrino detection energy threshold to about 5~GeV. IceCube includes a
surface cosmic ray air shower detector,
IceTop~\cite{IceCube:2012nn}. IceCube operates continuously through the year with
uptime over 99\%.

IceCube DOMs detect light from particle interactions in the ice. The energy,
position, time, and direction of the interacting particles are reconstructed from
the pattern of light deposition in the DOMs~\cite{Aartsen:2013vja}. Most
IceCube events at trigger level are downgoing muons from cosmic ray
air showers in the southern sky atmosphere, observed at rates of
2500--2900~Hz. Neutrinos from cosmic ray air showers are observed at
the rate of a few mHz. The atmospheric neutrino background consists of the  ``conventional'' background from the decay of pions and kaons and the harder spectrum but much less abundant ``prompt'' background from the decay of charmed mesons. Particle interaction topologies in IceCube are
flavor- and interaction-dependent and fall into two
primary categories: linear ``tracks'' and quasi-spherical
``showers''. Tracks result from muons, either from cosmic ray
backgrounds or from \numu charged-current (CC) interactions. Showers
arise from neutral current (NC) interactions of all neutrino flavors,
and from CC interactions of \nue and most \nutau. A small fraction of
high-energy \nutau may produce a double cascade from the initial CC 
neutrino interaction and subsequent tau lepton decay. Nearly all high energy neutrino interactions, regardless of flavor, are deep inelastic scattering off nucleons in the ice. However, electron anti-neutrinos can also interact with electrons in the ice via the Glashow resonance, with interaction probability sharply peaked at 6.3~PeV.

IceCube was designed to detect astrophysical neutrinos from potential cosmic ray acceleration sites. Assuming that charged cosmic rays are accelerated via the Fermi mechanism in astrophysical environments with strong shocks, the expected astrophysical neutrino flux is 10 -- 100 events per year in IceCube above the atmospheric background, with a power law energy distribution close to $E^{-2}$. Neutrino production in cosmic ray acceleration sites has a natural application in multi-messenger astronomy since charged cosmic rays and gamma rays are expected to interact with protons and photons at the site to produce pions. Charged pion decays produce neutrinos with a flavor ratio of 1:2:0 \nue:\numu:\nutau (treating neutrinos and anti-neutrinos as identical). Neutral pion decays produce gamma rays. Therefore, some gamma ray sources should also be neutrino emitters.
%

IceCube's science program also covers neutrino oscillation physics
down to 5~GeV with IceCube DeepCore, indirect
dark matter detection, sterile neutrino searches, searches for other physics beyond the Standard Model
such as Lorentz invariance violation, cosmic ray physics, and
glaciological studies of the South Pole ice.

This talk reviews recent results from IceCube, excluding cosmic ray measurements, which are discussed in~\cite{soldinpos2019}.

\section{Diffuse Astrophysical Neutrino Flux}
IceCube uses two
methods to separate astrophysical neutrinos from the atmospheric
background. One method selects through-going track-like events from the northern
hemisphere. These tracks originate from outside the instrumented
volume, increasing the effective volume for neutrino detection. The
Earth filters out muons from the northern hemisphere, and energy is
used to discriminate the expected hard ($E^{-2}$) astrophysical
neutrino component from the soft atmospheric neutrino spectrum. However, this method is not sensitive to the southern hemisphere sky, nor is it sensitive to cascade-type events
which should account for the majority of the astrophysical neutrino
flux, since standard neutrino flavor oscillation should result in approximately a 1:1:1
ratio of \nue:\numu:\nutau at the Earth. The other method selects
events of both cascade and track type from the whole sky, but
requires that the events start inside the detector, in order to
eliminate through-going atmospheric muons which should deposit light on the outer strings of the detector. Tracks in the starting event sample may exit the detector, and therefore are not fully contained. Most cascades in this sample are fully contained. 

A new high energy cascade sample is being developed which does not require containment. The highest energy event from this sample is the first Glashow resonance candidate observed by IceCube~\cite{lulupos2017}. The sample is being extended with updated reconstruction which uses detailed timing information in order to separate high energy cosmic ray muons from partially contained neutrino cascade events~\cite{lulupos2019}.

IceCube announced the first detection of a diffuse flux of
cosmic neutrinos in 2013, using the high-energy all-flavor starting event
sample from the whole sky~\cite{Aartsen:2014gkd}. The discovery was confirmed by the
high-energy through-going track sample from the northern hemisphere
sky~\cite{Aartsen:2016xlq}. In both samples the arrival directions of the neutrinos were
consistent with an isotropic distribution. 
\subsection{High Energy Starting Events}
The high energy starting event sample consists of track and cascade events with energies above 60~TeV. This sample has been updated to include 7.5~years of data from 2010 to 2017~\cite{hesepos2019}. Treatment of atmospheric background components has also been updated. Analyses of earlier iterations of this sample allowed the normalization of background components to float. The updated analysis takes into account the pion/kaon ratio and neutrino/anti-neutrino fraction in the conventional atmospheric neutrino flux, and the spectral index of the primary cosmic rays. Considering a single power law fit to the data $\frac{d\Phi_{\nu+\bar\nu}}{dE}=\Phi {\left(\frac{E_\nu}{\textrm{100 TeV}}\right)}^{-\gamma} \cdot 10^{-18}~[\textmd{GeV}^{-1}\textmd{cm}^{-2}\textmd{s}^{-1}\textmd{sr}^{-1}]$, the best fit spectral index $\gamma$ is $2.89^{+0.20}_{-0.19}$ with an all-flavor flux normalization $\Phi$ of $6.45^{+1.46}_{-0.46}$. The data does not prefer a broken power law model. 
\subsection{High energy Through-going Tracks}
\begin{figure*}
    \centering
    \includegraphics[width=0.95\linewidth]{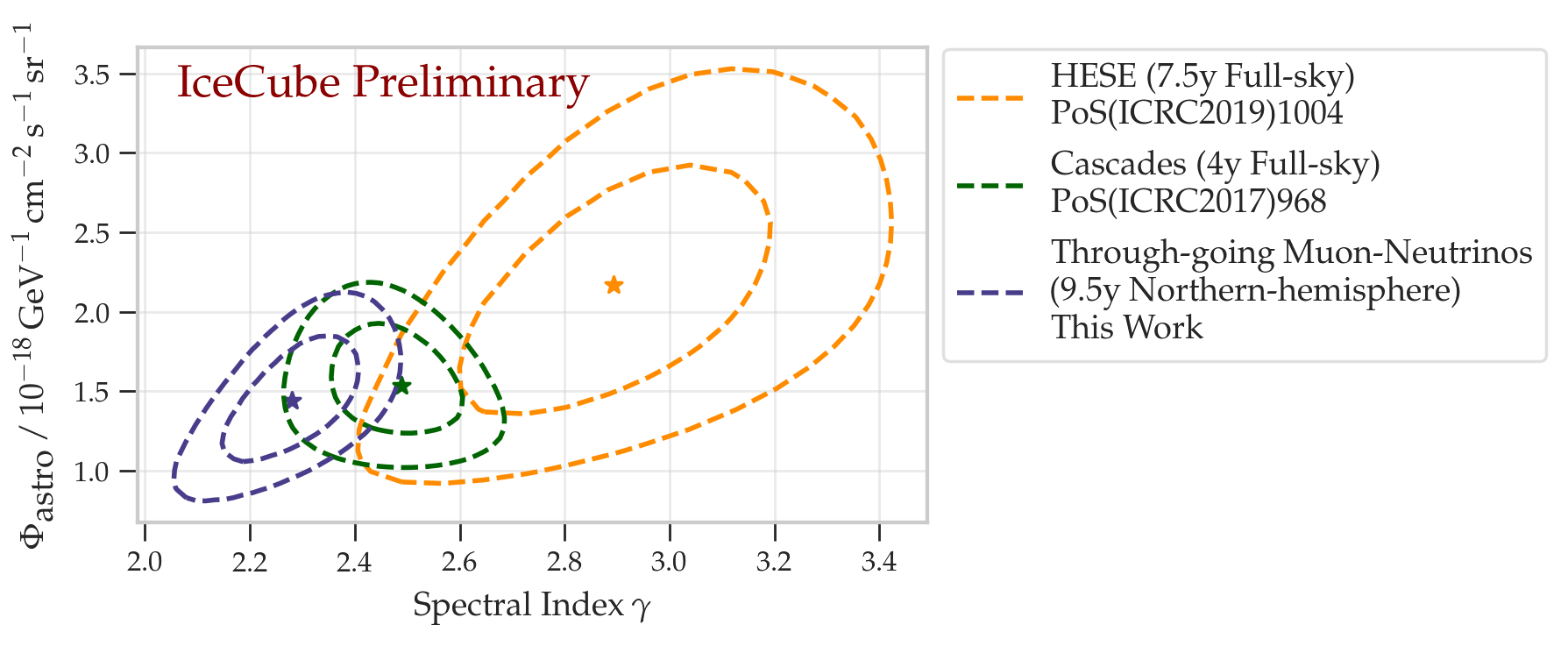}
    \caption{Best fit flux normalization and spectral index for a single power law fit to through-going tracks (blue), contained cascades (green) and starting tracks + cascades (yellow). Inner/outer contours are 68\% and 99\% uncertainties respectively. Note that the all-flavor flux normalization of the starting event sample is divided by 3 to compare to single-flavor normalizations.}\label{fig:diffuse}
\end{figure*}

The through-going track sample has been updated to use nearly 10~years of data from May 2009 to December 2018~\cite{stettnerpos2019}. The best fit single power law spectral index is $2.28^{+0.08}_{-0.09}$, and the best fit single flavor flux normalization is $1.44^{+0.25}_{-0.24}\cdot 10^{-18}~[\textmd{GeV}^{-1}\textmd{cm}^{-2}\textmd{s}^{-1}\textmd{sr}^{-1}]$. The central energy range is 40~TeV to 3.5~PeV. This result is consistent with earlier iterations of the sample, although the spectral index is now softer than previously reported. The best fit is also consistent with an independent sample of starting events designed to study the inelasticity distribution~\cite{Aartsen:2018vez}. Figure~\ref{fig:diffuse} shows the best fit flux normalization and spectral index for the through-going track sample, the starting event sample, and also the sample of contained cascades presented in 2017~\cite{cascade2017}. There is no evidence of the prompt atmospheric neutrino component in the sample; work to place an upper limit on the prompt neutrino flux is ongoing. A global fit of the diffuse samples is forthcoming.

\subsection{Tau Neutrino Searches}
Assuming that the flavor ratio at the source is 1:2:0::\nue:\numu:\nutau, standard neutrino oscillation physics predicts a flavor ratio of 1:1:1 in IceCube. Even in extreme cases such as pure \nue or pure \numu composition at the source, the flavor ratio in IceCube will still include a significant fraction of \nutau after standard oscillations. Several analyses of the flavor ratio in IceCube have been performed~\cite{Aartsen:2018vez,Aartsen:2015ivb,Aartsen:2015knd} but a limiting factor has been the lack of clearly identified tau neutrino candidates. Most \nutau appear as single cascades in IceCube, and therefore the experimental signal is degenerate with that of neutral current events and \nue charged current events. 
The signature of a high energy \nutau charged current event is a cascade from the neutrino deep inelastic scattering interaction, followed by a signature from the decay of the tau lepton. The tau lepton decays 65\% of the time to hadrons and 18\% to electrons. In both cases a second cascade is seen, producing a double cascade signature~\cite{Learned:1994wg}. The remaining decays are to muons; this decay mode is not considered here. 
In some cases, even if the event appears as a single cascade, the two cascades may be distinguished as double pulses in individual IceCube DOMs, thanks to the digitized PMT signal (waveform) captured by the DOM. A search for double pulses was published by IceCube using three years of data~\cite{Aartsen:2015dlt} with negative results.

\begin{figure*}
    \centering
    \includegraphics[width=0.48\linewidth]{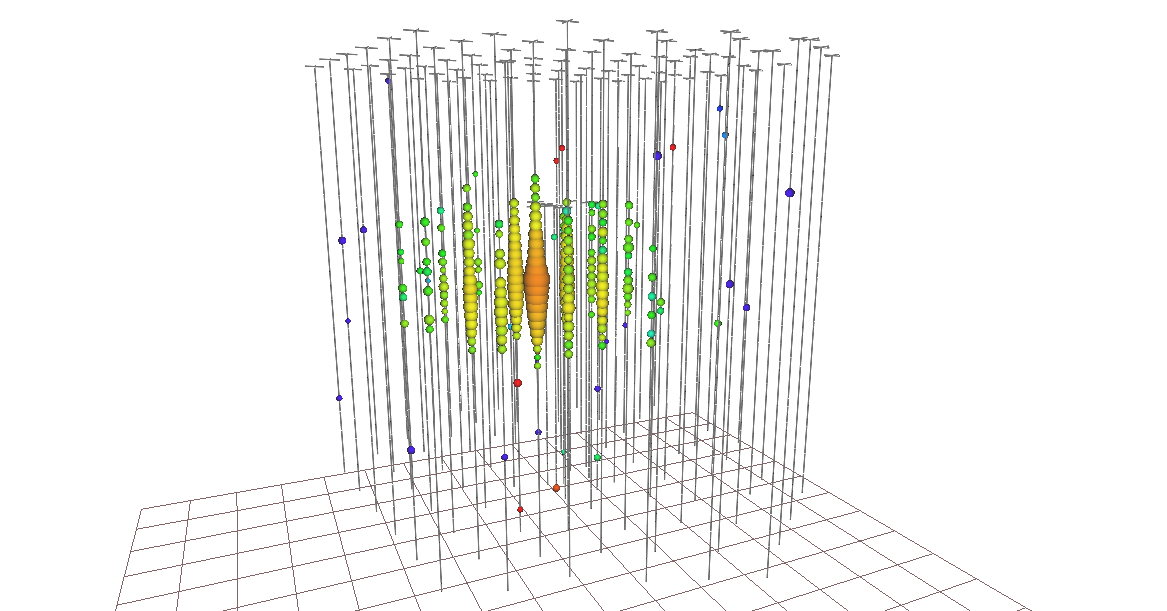}
    \includegraphics[width=0.48\linewidth]{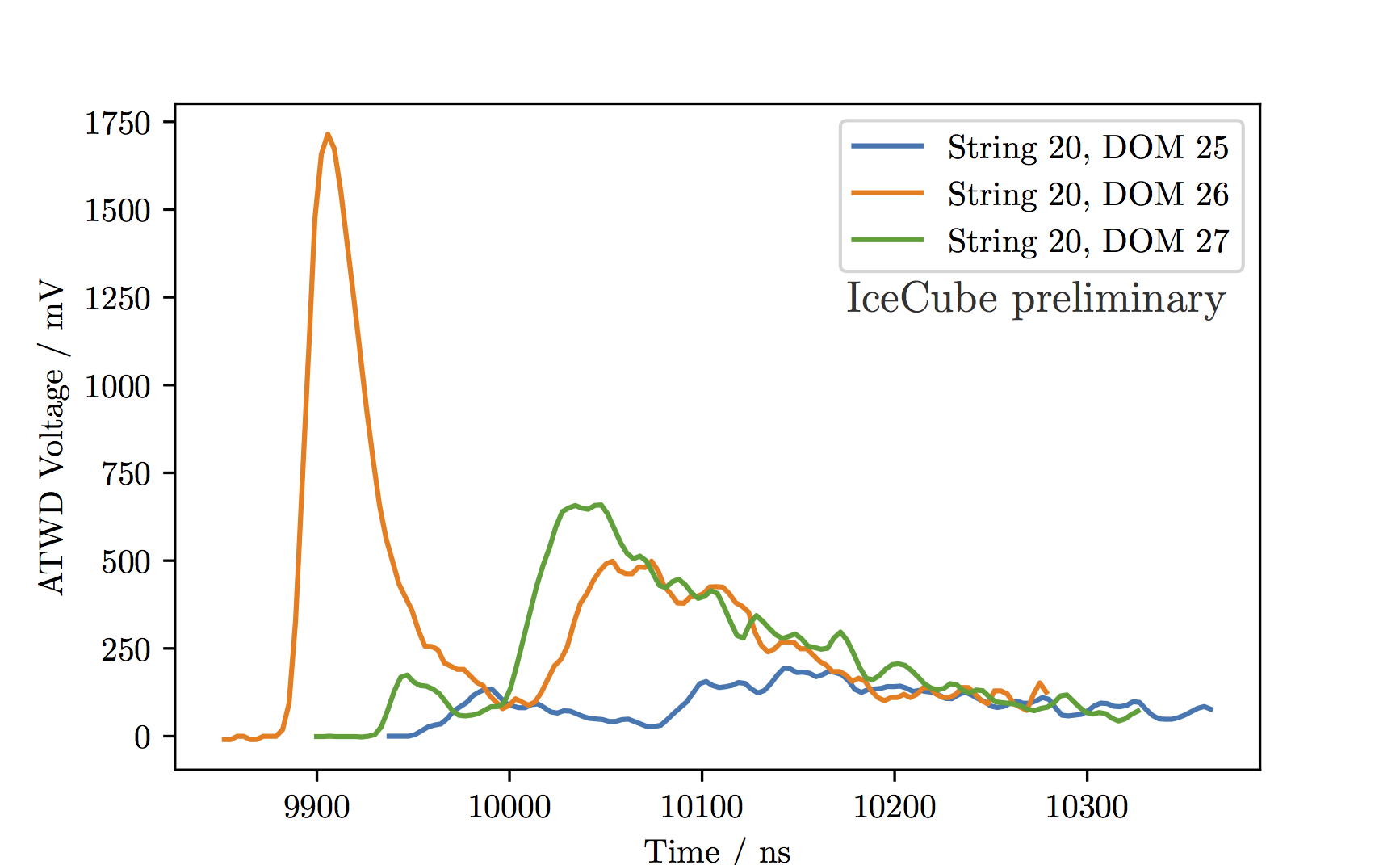}
    \caption{Left: event view of the only event selected by all three tau neutrino searches. Right: double pulse waveforms in the event. The orange waveform passes the first published double pulse search. Both the orange and green waveforms pass the updated double pulse search. All three waveforms pass the machine learning double pulse search.}\label{fig:taudp}
\end{figure*}

\begin{figure*}
    \centering
    \includegraphics[width=0.80\linewidth]{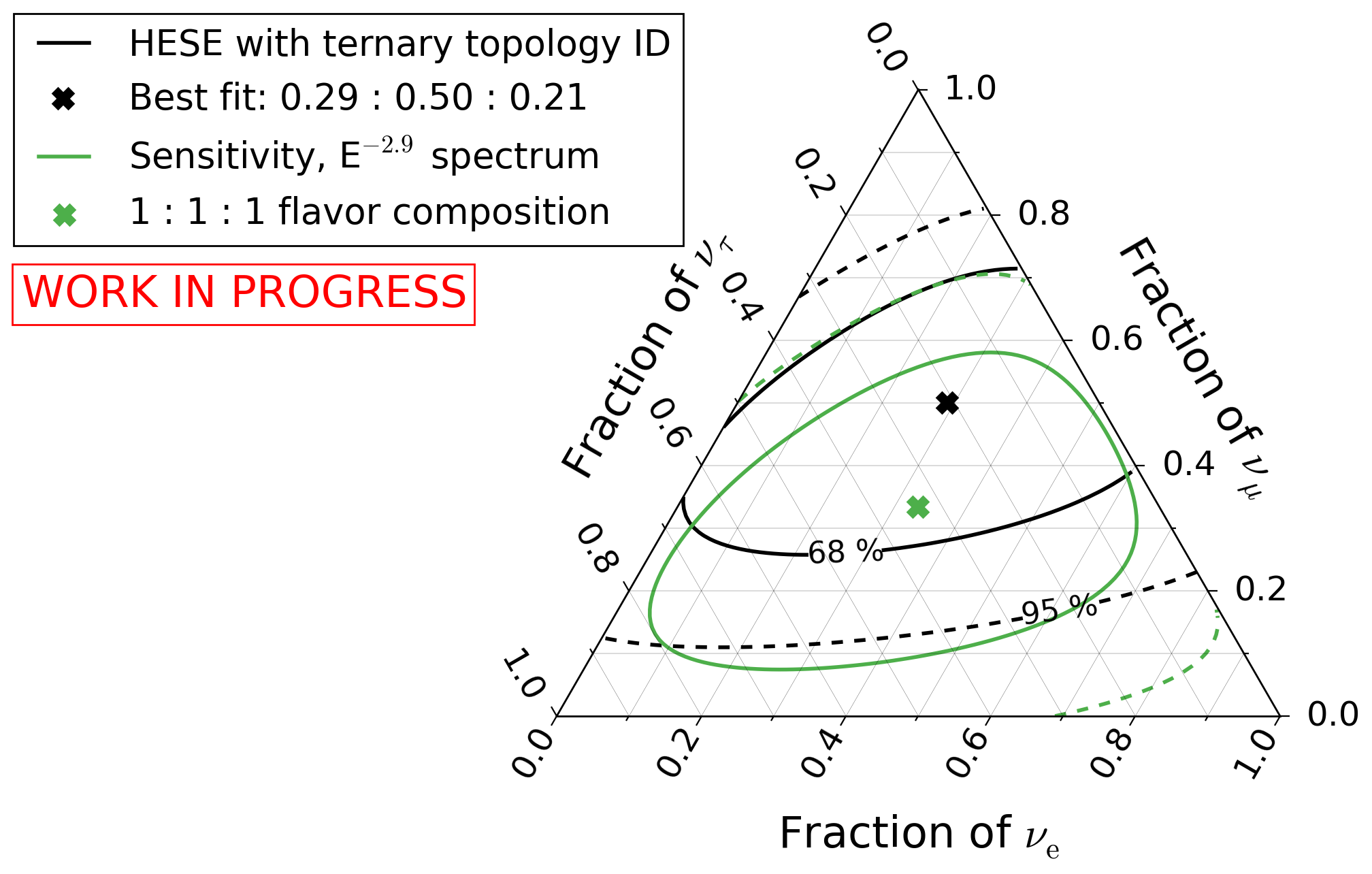}
    \caption{Sensitivity to flavor ratio (green) and results from the search using the double cascade tau identifier in the 7.5~year starting event sample.}\label{fig:flavor}
\end{figure*}

The 7.5~year starting event sample includes for the first time a double cascade topology identifier~\cite{juliannapos2019}. Each event is fitted to a double cascade likelihood hypothesis and double cascade candidates are selected based on the event length, the asymmetry between the two fitted cascade energies, and the fraction of the total energy deposited close to the fitted cascade vertices. Individual waveforms are not used in this search. In parallel, two updated double pulse searches have been performed: one~\cite{xupos2019} updates the double pulse method published in~\cite{Aartsen:2015dlt}, and a second which uses machine learning techniques~\cite{meierpos2019}. Both of these methods use individual DOM waveforms as well as looking at the overall shape of the event and other variables to eliminate background. Only one event passes all three searches. This event is shown in Figure~\ref{fig:taudp}, along with the double pulse waveforms. By eye, this event appears as a single cascade with a deposited energy of 89~TeV, but it has a double cascade event topology with a deposited energy of 9~TeV in the first cascade and 80~TeV in the second, and a distance of 17~m between the cascades. The best fit \nue:\numu:\nutau flavor ratio to the starting event sample, using the tau double cascade identifier, is 0.29:0.5:0.21, shown in Figure~\ref{fig:flavor}.  

\subsection{High energy Neutrino Physics}
The large high energy neutrino sample collected by IceCube enables neutrino physics studies at energies many times higher than at accelerators. The diameter of the Earth is equal to one absorption length for neutrinos at 40~TeV, therefore the neutrino-nucleon interaction cross section can be measured in IceCube above that energy by measuring the flux of up-going neutrinos as a function of energy and zenith angle. By contrast, this cross section has been measured at accelerators only up to 400~GeV. In 2018 IceCube published a cross section measurement from 6.3~TeV to 980~TeV, using one year of track events~\cite{Aartsen:2017kpd}. The result was consistent with the prediction of the Standard Model at next-to-leading order, using the parton distribution functions measured at HERA~\cite{CooperSarkar:2011pa}. An extension of this measurement to 8~years of data is in progress~\cite{sallypos2019}.

The neutrino interaction cross section has also been measured from 60~TeV to 10~PeV using the 7.5~year high energy starting event sample, which includes both tracks and cascades from the entire sky. The cross section measurement from this sample~\cite{tianlupos2019} is also consistent with the Standard Model.

Another diagnostic in starting events is the inelasticity, which describes the ratio of the energy in the hadronic cascade from the neutrino-nucleon interaction to total neutrino energy. IceCube published a measurement of the inelasticity from 1~TeV to 100~TeV in starting events in 5 years of data~\cite{Aartsen:2018vez}, the observed distribution as a function of energy is consistent with the same Standard Model calculation~\cite{CooperSarkar:2011pa}.

\section{Neutrino Sources}
The isotropic distribution of high energy neutrinos indicates that at least
part of the flux is extragalactic in origin. Searches for neutrino
excesses in IceCube have shown that the Galactic plane contributes $< 14\%$~\cite{Aartsen:2017ujz} of the observed neutrino flux.

The IceCube data sample has been searched multiple times for clustering that would indicate a point source or sources~\cite{Aartsen:2016oji,Aartsen:2018ywr}. Searches for neutrino point sources in IceCube have used both the northern hemisphere through-going track sample described above and a sample of through-going tracks from both hemispheres which is optimized for point source searches. A unified sample using 10 years of data has been generated from both track samples~\cite{tessapos2019}. The point source search uses a maximum likelihood ratio to compare the hypothesis of point-like signal plus isotropic background to an isotropic background-only null hypothesis. Additionally, an updated catalog of 110~sources was created, using gamma ray data to select gamma-bright sources that may produce neutrinos. The size of the catalog is chosen so as to limit trial factors applied to the most significant source in the catalog. Catalog objects include active galactic nuclei including blazars, starburst galaxies and Galactic gamma ray sources. The brightest neutrino source in the Northern Hemisphere sky coincides with the brightest catalog source, NGC~1068 (M77), a Seyfert II galaxy located at a distance of 14.4~Mpc. The significance of the source is 2.9~$\sigma$ after accounting for trials. The entire Northern source catalog is inconsistent with the isotropic background only hypothesis at 3.3~$\sigma$ due to excesses in the directions of NGC 1068, and the blazars TXS 0506+056, PKS 1424+240, and GB6 J1542+6129. When removing \txs, the excess is 2.25~$\sigma$. The brightest source in the southern hemisphere has a 55\% post trial significance, fully compatible with background.~\cite{tessapos2019}

\begin{figure*}
    \centering
    \includegraphics[width=0.49\linewidth]{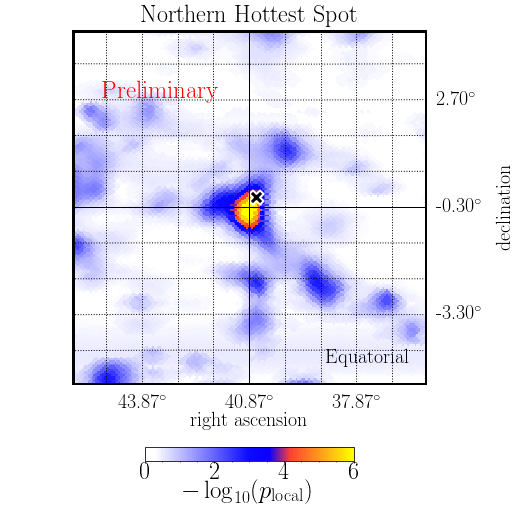}
    \includegraphics[width=0.49\linewidth]{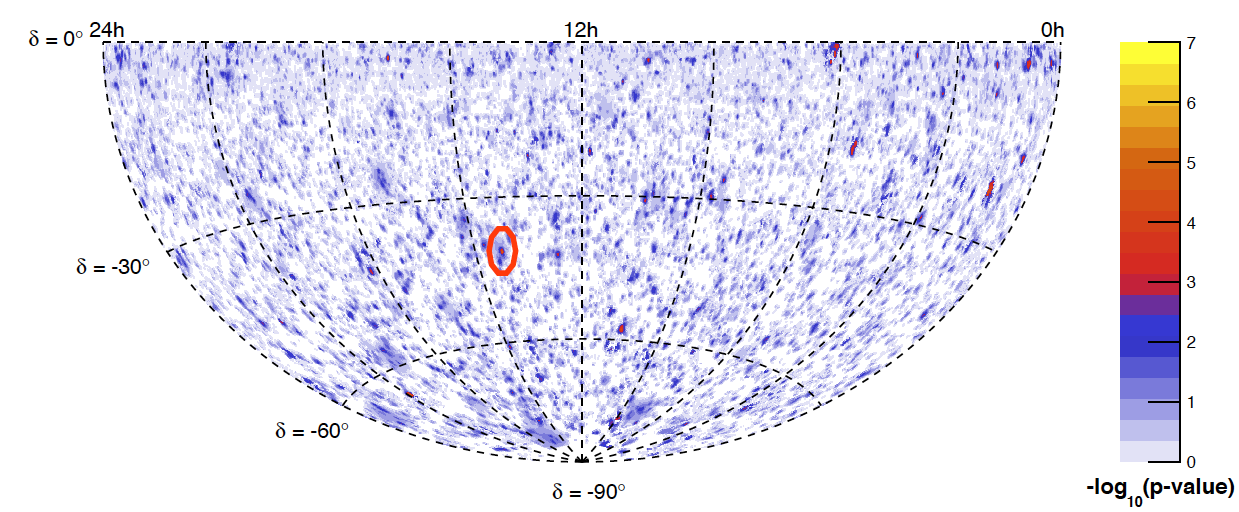}
    \caption{Left: Neutrino sky map near NGC 1068 using the 10 year IceCube combined track sample. Right: Southern Sky neutrino map with the combined IceCube and ANTARES search. The most signficant cluster is marked in red.}\label{fig:pointsource}
\end{figure*}

IceCube's location at the South Pole limits its sensitivity to the low energy neutrino sky in the southern hemisphere, due to the strong cuts necessary to reduce cosmic ray muon background from the southern sky. IceCube has partnered with ANTARES, an underwater neutrino telescope in the northern hemisphere, in order to do a joint point source search of the whole sky with maximum sensitivity~\cite{jointpspos2019}. The combined IceCube and ANTARES data sample uses 9 years of track and cascade data from ANTARES and 7 years of through-going track data from IceCube. The combined sensitivity to sources in the southern sky is a factor of 2 better than individual searches. A point source search was performed in the southern sky using angular extensions of 0\mydeg, 0.5\mydeg, 1\mydeg and 2\mydeg. The most significant source post-trial has a p-value of 18\% (0.9~$\sigma$). The Galactic center region was also searched with the same angular extensions within an ellipse 15\mydeg in Galactic latitude and 20\mydeg in Galactic longitude. The most significant spot in this region has a post-trials p-value of 3\% (1.9~$\sigma$). A catalog of 57 events was searched for neutrino clusters; the most significant source was HESSJ1023-575, a TeV gamma ray source associated with a young stellar cluster; the significance post trial was 42\% (0.2~$\sigma$). No significant results were seen from dedicated searches of Sagittarius A* or the shell type supernova remnant RXJ 1713.7-3946. Figure~\ref{fig:pointsource} shows the neutrino cluster seen by IceCube near NGC~1068 and the neutrino map from the southern sky combined analysis. Although no significant results were seen from the combined search, the analysis demonstrates the potential of combining complementary information from neutrino telescopes. IceCube and ANTARES also published joint limits on neutrino emission from the Galactic plane in 2018~\cite{Albert:2018vxw}.

IceCube is improving its individual source sensitivity in the southern sky by using contained cascades~\cite{Aartsen:2017eiu} and by developing a new lower-energy down-going starting track selection~\cite{estespos2019}. The new selection utilizes the fact that IceCube's triangular grid geometry increases the probability of background events passing the veto region in certain directions. Reconstruction of track events in multiple directions is used to determine the probability of any track being a background event. The resulting sample has a high astrophysical purity with less than 1 muon passing per year; the sample is most sensitive at declinations south of -30\mydeg at energies between 8 and 200~TeV. Lower energy starting tracks are naturally more sensitive to softer sources, and cascades are more sensitive to extended sources, both of which characterize the Galactic plane.

\subsection{Multi-messenger Astronomy with Neutrinos}
IceCube observes the entire sky continuously, and archives all data going back to the construction phase of the detector. IceCube is therefore well positioned to participate in the burgeoning field of multi-messenger astronomy, both in offline time-integrated and in real-time campaigns.

High energy neutrinos are expected to originate from cosmic ray acceleration sites. IceCube joined with ANTARES, the Pierre Auger Observatory (PAO) and the Telescope Array (TA) to search for correlations between the arrival directions of neutrinos and ultra high energy cosmic rays (UHECR)~\cite{uhecrpos2019}. Since cosmic rays are charged, they are deflected in flight by magnetic fields; since the deflection decreases with energy, the UHECR samples are limited to $\geq 57$~EeV in TA and $\geq 52$~EeV in PAO. The average deflection for 100~EeV UHECR is calculated to be 2.4\mydeg in the northern hemisphere and 3.6\mydeg in the southern hemisphere for a pure proton composition; the deflection will increase for heavier compositions, and is dependent on the chosen Galactic magnetic field model. Three analyses are performed: a cross-correlation between neutrino and UHECR directions with separation angle from 1\mydeg to 30\mydeg in 1\mydeg increments, a stacked likelihood search for clusters of UHECR events in the arrival direction of neutrino events, and a stacked likelihood search for clusters of neutrino events in the arrival direction of UHECR events.

The results from the analyses do not allow any conclusions about correlation between the arrival direction of neutrinos and UHECR. The strongest correlation in the cross-correlation analysis is at a separation angle of 14\mydeg for tracks and 16\mydeg for cascades with trials-corrected p-values of 23\% and 15\% respectively. The neutrino stacking analysis shows under-fluctuations for tracks, and for cascades the strongest result is at 3 times the average expected deflection of UHECR, with a corrected p value of 90\%. The UHECR stacking analysis shows the strongest result at a deflection of 6\mydeg for UHECR energies $\geq 85$~EeV. Interpretation of null results is not straightforward due to natural limitations of the analysis: TeV neutrinos may not come from the same sources as EeV cosmic rays, and EeV cosmic rays are expected to originate within 10-100 Mpc depending on composition whereas neutrinos can propagate over cosmological distances. 

In contrast to charged cosmic rays, gamma rays point back to their sources. IceCube has engaged in extensive multi-messenger campaigns with gamma ray telescopes. The High Altitude Water Cherenkov (HAWC) gamma ray observatory has a unique synergy with IceCube, since its field of view is in the northern hemisphere where IceCube's sensitivity to point sources at TeV energies is greatest. HAWC observes gamma rays from 0.3~TeV to 100~TeV in energy, which matches the energy of IceCube's point source neutrino sample well. IceCube has conducted a joint search of the Galactic plane in the northern hemisphere using a sky map of 1128 days of HAWC data and 8 years of neutrino tracks from the northern hemisphere sky. One search is a stacked search of 20 HAWC sources which are not identified with pulsar wind nebulae, and there are also 4 template searches of the northern Galactic plane, the Cygnus region, and the regions around J1908+063 and J1857+027. The most significant correlation is in the J1857+027 region, the p-value is 2\% before accounting for trials. Upper limits have been set on neutrino emission from all of these sources~\cite{hawcpos2019}.

The Antarctic Impulsive Transient Antenna (ANITA) is a balloon-borne experiment looking for Askaryan radio emission from neutrinos interacting in South Pole ice. ANITA is optimized for the discovery of cosmogenic neutrinos arising from UHECR interacting with the cosmic microwave background. ANITA has published three neutrino candidates, of which one is a cosmogenic neutrino candidate, whereas the other two are \nutau candidates~\cite{Allison:2018cxu,Gorham:2016zah,Gorham:2018ydl}. A \nutau signal would come from the neutrino interacting under the surface of the ice and the exiting tau lepton producing an up-going air shower detected by ANITA. The neutrino interpretation of the ANITA events is not consistent with the IceCube diffuse flux, but the neutrinos could come from point sources. IceCube searched for neutrinos from the direction of the ANITA events, looking for time-integrated excess of neutrinos as well as neutrino flares both in and out of temporal coincidence with the ANITA event times. No significant correlation was seen between the ANITA events and the IceCube data sample. Constraining limits can be set  on the secondary neutrino flux that would arise from \nutau neutrinos of these energies traversing the Earth.

\subsection{Realtime Multi-messenger Astronomy with IceCube}
A key element of multi-messenger astronomy is rapid communication of interesting sources to the observing community in order to locate counterparts in different messengers or wavelengths.  Following the discovery of high energy astrophysical neutrinos by IceCube, the collaboration implemented low-latency public alerts triggered by high energy neutrino track events~\cite{Aartsen:2016lmt}. Both high energy through-going and starting tracks are used to generate alerts. Cascades are not used at present due to their poor angular resolution, but they may be added to the alert stream in the future. The lower energy starting tracks mentioned above will also be added to the alert stream in the future. IceCube has transmitted public alerts since April 2016 using the Gamma-ray Coordination Network (GCN). The initial alert is transmitted with a median latency of 33~s following the neutrino interaction in the detector. The initial alert is issued via the GCN Notice, a machine-readable format designed to be generated and parsed quickly. A more time-consuming direction and energy reconstruction is performed on the alert event offline and a human-readable, citeable GCN Circular is issued generally a few hours after the initial Notice. The Circular can be used to retract the event if necessary, for example if a coincident cosmic ray muon causes a mis-reconstruction or mis-identification of the event. 
%

After the first three years of real-time operations, the alert system was improved in order to increase the number and signal purity of neutrinos and to reduce the number of subsequently retracted events. The alert text was also updated for improved clarity. The updated alerts are divided into ``Gold'' and ``Bronze'' categories, corresponding to 50\% and 30\% signal probability respectively. The alert text includes the time and date of the event, the reconstructed right ascension and declination, the angular radii of the 50\% and 90\% containment circles and the signal probability. The new alerts have been functional since June 17, 2019. Gold alerts are expected to be issued at the rate of one per month, and bronze alerts at the rate of 1.3 per month~\cite{realtimepos2019}.

\subsection{Coincident Detection of a Flaring Blazar and a High-Energy Neutrino}

On September~22, 2017, IceCube sent an alert to GCN. The neutrino
which prompted the alert, denoted IC-170922A, was an upgoing,
through-going track event with a most likely energy of 290~TeV and a
55.6\% probability of being of astrophysical
origin~\cite{IceCube:2018dnn}. Following the alert, Fermi-LAT detected increased gamma ray flux from a known
blazar, \txs, which was located inside the directional uncertainty contour
of IC-170922A~\cite{IceCube:2018dnn}. The Major Atmospheric Gamma Imaging Cherenkov
Telescopes (MAGIC) then followed \txs~and detected
gamma rays at energies up to 400~GeV in 12~hours of observations
between September 24 and October 4~\cite{IceCube:2018dnn,Ahnen:2018mvi}. This was the first detection of
gamma rays at those energies from \txs.  Correlation between the gamma ray emission and
the high-energy neutrino is preferred over a chance coincidence at the
3$\sigma$ level~\cite{IceCube:2018dnn}. The redshift of the blazar was unknown prior
to the observation of IC-170922A. Following the multi-messenger observations of \txs, the
redshift was measured to be $0.3365 \pm 0.0010$~\cite{Paiano:2018qeq}.

\begin{figure*}
    \centering
    \includegraphics[width=0.98\linewidth]{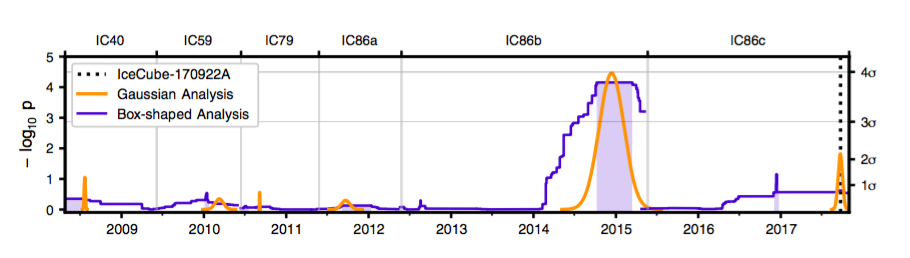}
    \caption{IceCube archival data from the region of
        IC-170922A, Apr 5, 2008 to Oct 31, 2017.}\label{fig:neutrinoflare}
\end{figure*}

IceCube archival data
was searched in the region of IC-170922A and an excess of high-energy neutrino events with respect to atmospheric
backgrounds was observed between September 2014 and March 2015~\cite{IceCube:2018cha}. The
best fit
Gaussian time window to the excess is centered on December 13, 2014,
with a duration of $110^{+35}_{-24}$~days. The observed excess is
$13\pm5$ events above the expected background from atmospheric
neutrinos. The excess is inconsistent with the background-only
hypothesis at the $3.5~\sigma$
level~\cite{IceCube:2018cha}. Figure~\ref{fig:neutrinoflare} shows the archival data
from this region in IceCube, and the best fit Gaussian time window to
the neutrino excess, along with results from a complementary box-shaped time window analysis.

\subsection{Sub-threshold Multi-messenger Searches with AMON}
IceCube participates in the Astrophysical Multi-messenger Observatory Network (AMON). AMON was designed to allow electromagnetic, neutrino, cosmic ray and gravitational wave observatories to share sub-threshold alerts with each other in order to develop coincident analyses. Events which are sub-threshold in any individual detector may be interesting if observed simultaneously in multiple detectors, in which case a public alert may be issued. 

IceCube has developed a sub-threshold coincident analysis with HAWC using AMON. IceCube sends high quality single tracks to AMON at the rate of 650 per day. Note that real-time alerts which are triggered by IceCube alone are at the much lower rate of one per month. HAWC sends source locations which show an excess over background when averaged over one transit of the source over the detector, at the rate of 800 per day. An initial study of 2 years of data (2016 and 2017) showed only one coincident event with an estimated false alarm rate of fewer than one per year (the FAR of this event is 1.1 per year). This analysis will form the basis of public alerts sent for coincidences with a FAR of 1 per year or lower~\cite{amonrealpos2019}.

\subsection{IceCube Response to External Events}
IceCube has developed a framework for rapid searches for neutrino emission from interesting transient events~\cite{fastfollowpos2019}. This can include flares from known sources such as the Crab Nebula or unknown bright transients. Additionally, all IceCube alert events are analyzed using time windows of $\pm$ 1 day and -30 days +1 day. In these cases the neutrino that triggered the alert is removed from the analysis. No statistically significant detections have been made using the IceCube fast response pipeline. Four follow-ups have yielded p-values below 0.035: PKS 0346-27 (a flaring blazar), the bright transient AT2018cow, IceCube alert track IC-180908A, and Fermi J1153-1124 (a flaring blazar).


\subsection{IceCube Analysis of Gravitational Wave Events}
The observation of a binary black hole merger by LIGO heralded the arrival of gravitational waves as astronomical messengers~\cite{Abbott:2016blz}. Gravitational waves joined the field of multi-messenger astronomy when a binary neutron star merger GW170817 was observed in coincidence with the short gamma ray burst GRB~170817A emitted by the kilonova resulting from the merger~\cite{GBM:2017lvd}. A massive multi-wavelength campaign was triggered by this event.

IceCube, along with ANTARES and PAO, searched for high energy neutrinos from the direction of the neutron star merger~\cite{ANTARES:2017bia}. No neutrinos were observed, possibly because the jet of the kilonova was not pointed directly at the Earth. IceCube also searched for neutrinos from all binary black hole mergers with null results. 

In the current O3 observing run of the LIGO/Virgo Collaboration (LVC) which began in April 2019, gravitational wave alerts are sent publicly. The expected alert rate is one binary black hole merger per week and one binary neutron star merger per month. IceCube uses two analyses to search for neutrinos from GW alert events~\cite{raamisgwpos2019,azadehgwpos2019}. One analysis is a maximum likelihood search for neutrino point sources using the gravitational wave map from LVC as a spatial prior, with a coincidence time window of $\pm500$~s. The other analysis is a Bayesian search considering the joint significance of GW and neutrino events with astrophysical priors such as the distance to the GW event. A GCN circular is sent by IceCube within about an hour of the alert from LVC. In the future, this process will be automated and the information sent as a GCN notice. Other future analysis plans include longer timescale searches for neutrinos produced later in the development of a kilonova and stacked analyses of the GW events. IceCube has also developed a GeV neutrino sample and used it to search for statistically significant increases around compact binary merger events. No significant excesses have been seen so far~\cite{gwenpos2019}.

\section{Beyond the Standard Model}
The nature of dark matter remains one of the outstanding problems in physics and astronomy. Assuming that dark matter consists of as yet unidentified non-relativistic weakly interacting massive particles expected from physics beyond the Standard Model, IceCube can search for dark matter indirectly by looking for anomalous neutrino signals that are inconsistent with astrophysical or atmospheric background. The most common search modes are for neutrinos arising from the decay or annihilation of dark matter into Standard Model particles; IceCube has published limits on dark matter interactions in the Sun~\cite{Aartsen:2016zhm}, the Earth~\cite{Aartsen:2016fep}, the Galactic center and halo~\cite{Aartsen:2017ulx,Aartsen:2016pfc,Aartsen:2018mxl}, and dwarf spheroidal galaxies~\cite{Aartsen:2013dxa}. 

IceCube has joined with ANTARES to search for dark matter annihilation in the Galactic center. IceCube and ANTARES have comparable sensitivity to the DM self-annihilation cross section in the Galactic center between 50~GeV and 1~TeV, so a combined analysis is more sensitive than the individual analyses of each detector. The combined analysis uses 3 years of tracks from IceCube and 9 years of tracks from ANTARES, and searches for signals of neutrinos from DM self-annihilation to $W^+W^-$, $\tau^+\tau^-$, $\mu^+\mu^{-}$ and $b\bar{b}$. No significant excess from the Galactic center was observed~\cite{antaresi3dmpos2019}. 

IceCube also searched for evidence of DM annihilation as well as DM decay from the Galactic center using the 7.5~year high energy starting event sample. In addition to the annihilation modes above, this analysis searched for annihilation to $\nu \bar{\nu}$. The decay search modes included all annihilation modes as well as $\nu_s \bar{\nu_s}$ and $H\nu$. In both cases the observational signature is an excess from the Galactic center above the astrophysical neutrino flux, but the signal from annihilation is more peaked toward the Galactic center than the signal from decay since annihilation probability depends quadratically on DM density whereas decay probability depends only linearly on DM density. No excess was observed, so limits were set in the 100~TeV to 10~PeV energy range on the dark matter lifetime and self-annihilation cross section~\cite{hesedmpos2019}.

A new search for neutrino scattering on dark matter was also performed with the 7.5~year starting event sample. In this case, the model is that the astrophysical neutrinos interact with DM in the Galactic center via scattering, and the observational signature is an anisotropy imprinted on the astrophysical neutrino flux. Two scattering models are considered: the DM is a scalar particle interacting with neutrinos via a fermionic mediator, and the DM is a fermionic particle interacting with neutrinos via a vector boson mediator. No anisotropy is observed in the astrophysical neutrino flux, so limits are set on the mediator mass. IceCube sets leading bounds on mediator masses above 10~MeV for DM masses between 1~MeV and 1~GeV~\cite{hesedmpos2019}. 

IceCube has published a number of other searches for beyond Standard Model physics including limits on sterile neutrinos~\cite{TheIceCube:2016oqi,Aartsen:2017bap}, monopoles~\cite{Aartsen:2015exf}, Lorentz violation~\cite{Aartsen:2017ibm} and non-standard neutrino interactions~\cite{Aartsen:2017xtt}. The agreement of the measured cross-section and, especially, the inelasticity with the Standard Model expectation also potentially constrains new physics such as leptoquarks and new dimensions at the electroweak scale. 

\section{Tau Neutrino Appearance from Atmospheric Neutrino Oscillations}
The densely instrumented DeepCore region in the center of IceCube is sensitive to neutrino events down to 5~GeV. The low energy sensitivity arises from the closer spacing of DeepCore strings, the higher quantum efficiency PMTs in DeepCore DOMs, and the exceptional optical properties of the ice in the bottom half of the detector. IceCube collects a very high statistics sample of atmospheric neutrinos which are used for neutrino oscillation studies. The probability of muon neutrino disappearance (and corresponding tau neutrino appearance) is maximal at 25~GeV for a propagation length equal to the diameter of the Earth. IceCube can access many other baselines and energies for neutrinos traveling at other angles through the Earth. IceCube's analysis of muon neutrino disappearance in three years of DeepCore data between 5 and 56~GeV had sensitivity to atmospheric neutrino oscillation parameters which is comparable to that of long baseline accelerator experiments~\cite{Aartsen:2017nmd}. IceCube's analysis is complementary to accelerator measurements since the neutrinos are much higher in energy and IceCube has very different detector systematic uncertainties to a long baseline experiment. 

Tau neutrino appearance is an important channel for probing the unitary of the PMNS neutrino mixing matrix; the tau sector is the most poorly constrained area of neutrino oscillations. Non-unitarity in the PMNS matrix could indicate the existence of sterile neutrinos or non-standard interactions of the three active neutrino flavors. Tau neutrinos at 5 -- 56~GeV energies do not show the double cascade topology of high energy astrophysical tau neutrinos; instead, the tau neutrinos appear as an excess of cascades compared to the expectation from no neutrino oscillation. The result is expressed as a tau neutrino normalization, the ratio of the number of tau neutrinos observed to the number of tau neutrinos expected based on standard oscillation parameters. IceCube's sensitivity to atmospheric \nutau appearance in 3 years of operation is comparable to results published by Super-K~\cite{Li:2017dbe} (atmospheric) and OPERA~\cite{Agafonova:2018auq} (accelerator). Both Super-K and OPERA measure a \nutau normalization slightly higher than 1. IceCube recently published results from a search for tau neutrino appearance in 3 years of data, measuring a \nutau normalization of $0.73^{+0.30}_{-0.24}$~\cite{Aartsen:2019tjl}. An updated analysis with 10~years data and updated reconstruction and treatment of systematic errors is in progress. This analysis should achieve a precision of 15\% in \nutau normalization.

\section{The IceCube Upgrade}
Although a neutrino has been observed in coincidence with the flaring blazar \txs, this source accounts for less than 1\% of the observed astrophysical neutrino flux. Excesses seen at the coordinates of 4 sources in the 10 year combined track sample, including NGC 1068 and \txs, are inconsistent with the background only hypothesis at 3.3~$\sigma$; more statistics are needed to confirm this result. Many of IceCube's highest energy neutrinos have no observed counterparts. The observation of astrophysical tau neutrinos and Glashow resonance events is at the limit of IceCube's sensitivity, with order of 1 event expected in 10 years. This motivates the development of a next generation multi-cubic-kilometer astrophysical neutrino detector, called IceCube-Gen2~\cite{Aartsen:2014njl}. 

An upgrade of the current IceCube detector has been approved~\cite{upgradepos2019}. The IceCube Upgrade will consist of 7 new strings to be deployed at the center of the detector inside the DeepCore region. The new strings will be instrumented with new sensors with increased photocathode coverage and ultraviolet sensitivity. The Upgrade will also include an extensive array of calibration devices in order to  measure the optical properties of the ice with unprecedented precision. The Upgrade will serve as a testing platform for some new device concepts including wavelength-shifting and optical fiber sensors which may be used in a future Gen2 detector. 

\begin{figure*}
    \centering
    \includegraphics[width=0.48\linewidth]{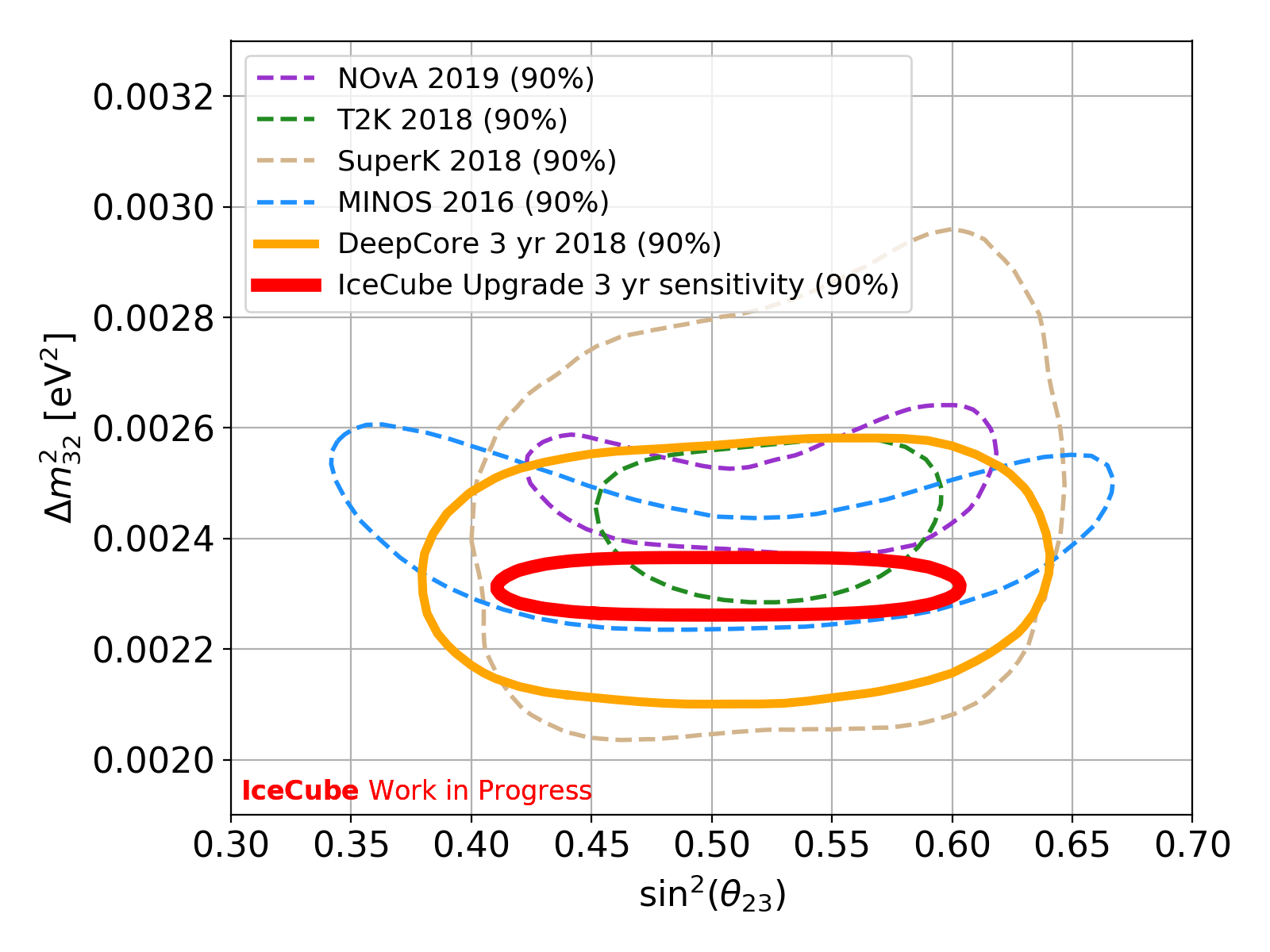}
    \includegraphics[width=0.48\linewidth]{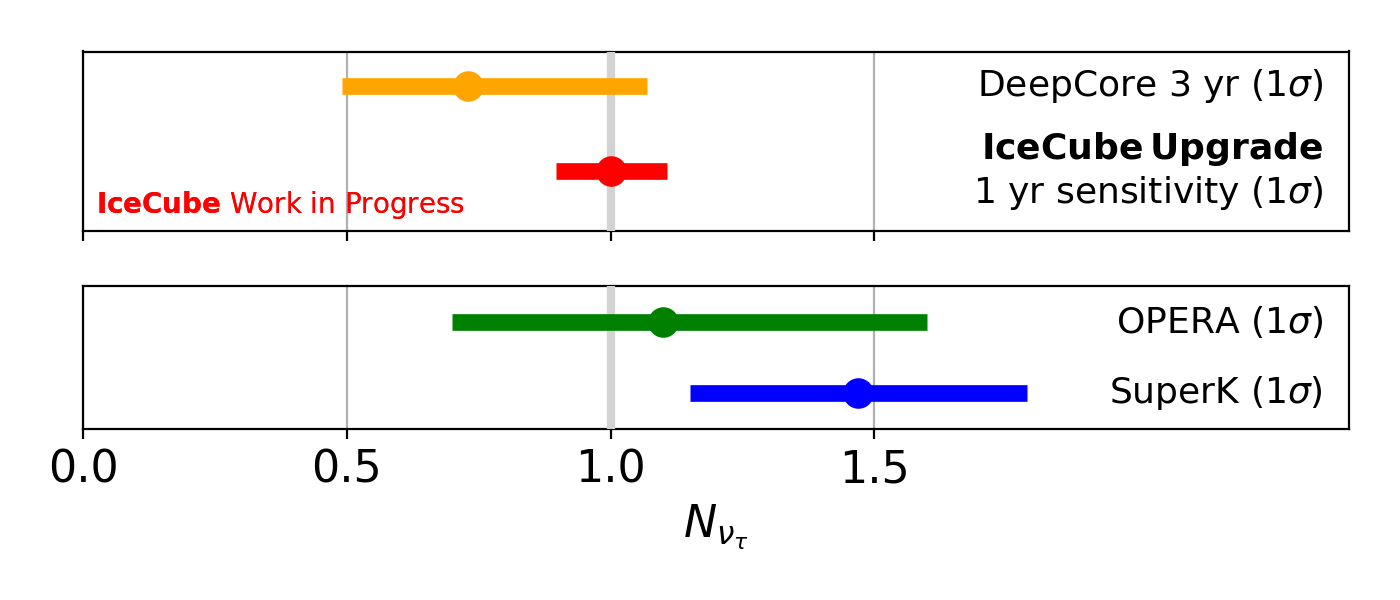}
    \caption{Left: projected sensitivity to atmospheric neutrino oscillation parameters in the upgrade. Right: projected sensitivity to \nutau neutrino normalization in the upgrade.}\label{fig:upgradeosc}
\end{figure*}

The Upgrade will collect 2 times more \nutau charged current interactions relative to DeepCore, with a factor of 3 improvement in resolution in the relevant energy range. The upgrade sensitivity to \numu disappearance will be comparable to the most sensitive existing accelerator experiments. More importantly, the upgrade will achieve better than 10\% sensitivity to \nutau normalization in one year of operations. The upgrade sensitivity to atmospheric neutrino oscillations is shown in Figure~\ref{fig:upgradeosc}.

The IceCube upgrade will also be sensitive to the dark matter self annihilation cross section for masses well below 10~GeV~\cite{upgradedmpos2019}. The updated calibration information will benefit both low energy analyses and astrophysical neutrino searches, since the calibration can be applied to the entire archival IceCube data set. 

\section{Conclusion}
IceCube plays a major role in the emerging field of multi-messenger astronomy, having observed the first correlation of an electromagnetic counterpart to a high energy cosmic neutrino source. IceCube pursues a broad science program and has published world leading results in beyond standard model physics, neutrino oscillations and high energy neutrino physics. An upgrade to the IceCube detector has been approved, and will produce new leading physics results and serve as the first step to the next generation neutrino telescope at the South Pole.





\bibliographystyle{ICRC}
\bibliography{references}

\providecommand{\href}[2]{#2}\begingroup\raggedright\begin{thebibliography}{10}

\bibitem{Aartsen:2016nxy}
{\bf IceCube} Collaboration, M.~G. Aartsen et~al., {\em JINST} {\bf 12} (2017)
  P03012.

\bibitem{Abbasi:2010vc}
{\bf IceCube} Collaboration, R.~Abbasi et~al., {\em Nucl. Instrum. Meth.} {\bf
  A618} (2010) 139--152.

\bibitem{Abbasi:2008aa}
{\bf IceCube} Collaboration, R.~Abbasi et~al., {\em Nucl. Instrum. Meth.} {\bf
  A601} (2009) 294--316.

\bibitem{Collaboration:2011ym}
{\bf IceCube} Collaboration, R.~Abbasi et~al., {\em Astropart. Phys.} {\bf 35}
  (2012) 615--624.

\bibitem{IceCube:2012nn}
{\bf IceCube} Collaboration, R.~Abbasi et~al., {\em Nucl. Instrum. Meth.} {\bf
  A700} (2013) 188--220.

\bibitem{Aartsen:2013vja}
{\bf IceCube} Collaboration, M.~G. Aartsen et~al., {\em JINST} {\bf 9} (2014)
  P03009.

\bibitem{soldinpos2019}
{\bf IceCube} Collaboration,  \pos{PoS(ICRC2019)014} (these proceedings).

\bibitem{lulupos2017}
{\bf IceCube} Collaboration,  \pos{PoS(ICRC2017)1002} (2017).

\bibitem{lulupos2019}
{\bf IceCube} Collaboration,  \pos{PoS(ICRC2019)945} (these proceedings).

\bibitem{Aartsen:2014gkd}
{\bf IceCube} Collaboration, M.~G. Aartsen et~al., {\em Phys. Rev. Lett.} {\bf
  113} (2014) 101101.

\bibitem{Aartsen:2016xlq}
{\bf IceCube} Collaboration, M.~G. Aartsen et~al., {\em Astrophys. J.} {\bf
  833} (2016) 3.

\bibitem{hesepos2019}
{\bf IceCube} Collaboration,  \pos{PoS(ICRC2019)1004} (these proceedings).

\bibitem{stettnerpos2019}
{\bf IceCube} Collaboration,  \pos{PoS(ICRC2019)1017} (these proceedings).

\bibitem{Aartsen:2018vez}
{\bf IceCube} Collaboration, M.~G. Aartsen et~al., {\em Phys. Rev.} {\bf D99}
  (2019) 032004.

\bibitem{cascade2017}
{\bf IceCube} Collaboration,  \pos{PoS(ICRC2017)968} (2017).

\bibitem{Aartsen:2015ivb}
{\bf IceCube} Collaboration, M.~G. Aartsen et~al., {\em Phys. Rev. Lett.} {\bf
  114} (2015) 171102.

\bibitem{Aartsen:2015knd}
{\bf IceCube} Collaboration, M.~G. Aartsen et~al., {\em Astrophys. J.} {\bf
  809} (2015) 98.

\bibitem{Learned:1994wg}
J.~G. Learned and S.~Pakvasa, {\em Astropart. Phys.} {\bf 3} (1995) 267--274.

\bibitem{Aartsen:2015dlt}
{\bf IceCube} Collaboration, M.~G. Aartsen et~al., {\em Phys. Rev.} {\bf D93}
  (2016) 022001.

\bibitem{juliannapos2019}
{\bf IceCube} Collaboration,  \pos{PoS(ICRC2019)1015} (these proceedings).

\bibitem{xupos2019}
{\bf IceCube} Collaboration,  \pos{PoS(ICRC2019)1036} (these proceedings).

\bibitem{meierpos2019}
{\bf IceCube} Collaboration,  \pos{PoS(ICRC2019)960} (these proceedings).

\bibitem{Aartsen:2017kpd}
{\bf IceCube} Collaboration, M.~G. Aartsen et~al., {\em Nature} {\bf 551}
  (2017) 596--600.

\bibitem{CooperSarkar:2011pa}
A.~Cooper-Sarkar, P.~Mertsch, and S.~Sarkar, {\em JHEP} {\bf 08} (2011) 042.

\bibitem{sallypos2019}
{\bf IceCube} Collaboration,  \pos{PoS(ICRC2019)990} (these proceedings).

\bibitem{tianlupos2019}
{\bf IceCube} Collaboration,  \pos{PoS(ICRC2019)1040} (these proceedings).

\bibitem{Aartsen:2017ujz}
{\bf IceCube} Collaboration, M.~G. Aartsen et~al., {\em Astrophys. J.} {\bf
  849} (2017) 67.

\bibitem{Aartsen:2016oji}
{\bf IceCube} Collaboration, M.~G. Aartsen et~al., {\em Astrophys. J.} {\bf
  835} (2017) 151.

\bibitem{Aartsen:2018ywr}
{\bf IceCube} Collaboration, M.~G. Aartsen et~al., {\em Eur. Phys. J.} {\bf
  C79} (2019) 234.

\bibitem{tessapos2019}
{\bf IceCube} Collaboration,  \pos{PoS(ICRC2019)851} (these proceedings).

\bibitem{jointpspos2019}
{\bf ANTARES,IceCube} Collaboration,  \pos{PoS(ICRC2019)919} (these
  proceedings).

\bibitem{Albert:2018vxw}
{\bf ANTARES, IceCube} Collaboration, A.~Albert et~al., {\em Astrophys. J.}
  {\bf 868} (2018) L20.

\bibitem{Aartsen:2017eiu}
{\bf IceCube} Collaboration, M.~G. Aartsen et~al., {\em Astrophys. J.} {\bf
  846} (2017) 136.

\bibitem{estespos2019}
{\bf IceCube} Collaboration,  \pos{PoS(ICRC2019)954} (these proceedings).

\bibitem{uhecrpos2019}
{\bf ANTARES, IceCube, Pierre Auger, Telescope Array} Collaboration,
  \pos{PoS(ICRC2019)842} (these proceedings).

\bibitem{hawcpos2019}
{\bf HAWC, IceCube} Collaboration,  \pos{PoS(ICRC2019)932} (these proceedings).

\bibitem{Allison:2018cxu}
{\bf ANITA} Collaboration, P.~W. Gorham et~al., {\em Phys. Rev.} {\bf D98}
  (2018) 022001.

\bibitem{Gorham:2016zah}
{\bf ANITA} Collaboration, P.~W. Gorham et~al., {\em Phys. Rev. Lett.} {\bf
  117} (2016) 071101.

\bibitem{Gorham:2018ydl}
{\bf ANITA} Collaboration, P.~W. Gorham et~al., {\em Phys. Rev. Lett.} {\bf
  121} (2018) 161102.

\bibitem{Aartsen:2016lmt}
{\bf IceCube} Collaboration, M.~G. Aartsen et~al., {\em Astropart. Phys.} {\bf
  92} (2017) 30--41.

\bibitem{realtimepos2019}
{\bf IceCube} Collaboration,  \pos{PoS(ICRC2019)1021} (these proceedings).

\bibitem{IceCube:2018dnn}
{\bf Liverpool Telescope, MAGIC, H.E.S.S., AGILE, Kiso, VLA/17B-403, INTEGRAL,
  Kapteyn, Subaru, HAWC, Fermi-LAT, ASAS-SN, VERITAS, Kanata, IceCube, Swift
  NuSTAR} Collaboration, M.~G. Aartsen et~al., {\em Science} {\bf 361} (2018)
  eaat1378.

\bibitem{Ahnen:2018mvi}
{\bf MAGIC} Collaboration, S.~Ansoldi et~al., {\em Astrophys. J. Lett.} (2018).
  [Astrophys. J.863,L10(2018)].

\bibitem{Paiano:2018qeq}
S.~Paiano, R.~Falomo, A.~Treves, and R.~Scarpa, {\em Astrophys. J.} {\bf 854}
  (2018) L32.

\bibitem{IceCube:2018cha}
{\bf IceCube} Collaboration, M.~G. Aartsen et~al., {\em Science} {\bf 361}
  (2018) 147--151.

\bibitem{amonrealpos2019}
{\bf HAWC, IceCube} Collaboration,  \pos{PoS(ICRC2019)841} (these proceedings).

\bibitem{fastfollowpos2019}
{\bf IceCube} Collaboration,  \pos{PoS(ICRC2019)1026} (these proceedings).

\bibitem{Abbott:2016blz}
{\bf LIGO Scientific, Virgo} Collaboration, B.~P. Abbott et~al., {\em Phys.
  Rev. Lett.} {\bf 116} (2016) 061102.

\bibitem{GBM:2017lvd}
{\bf LIGO Scientific,Virgo} Collaboration, B.~P. Abbott et~al., {\em Astrophys.
  J.} {\bf 848} (2017) L12.

\bibitem{ANTARES:2017bia}
{\bf Virgo, IceCube, Pierre Auger, ANTARES, LIGO Scientific} Collaboration,
  A.~Albert et~al., {\em Astrophys. J.} {\bf 850} (2017) L35.

\bibitem{raamisgwpos2019}
{\bf IceCube} Collaboration,  \pos{PoS(ICRC2019)918} (these proceedings).

\bibitem{azadehgwpos2019}
{\bf IceCube} Collaboration,  \pos{PoS(ICRC2019)930} (these proceedings).

\bibitem{gwenpos2019}
{\bf IceCube} Collaboration,  \pos{PoS(ICRC2019)865} (these proceedings).

\bibitem{Aartsen:2016zhm}
{\bf IceCube} Collaboration, M.~G. Aartsen et~al., {\em Eur. Phys. J.} {\bf
  C77} (2017) 146. [Erratum: Eur. Phys. J.C79,no.3,214(2019)].

\bibitem{Aartsen:2016fep}
{\bf IceCube} Collaboration, M.~G. Aartsen et~al., {\em Eur. Phys. J.} {\bf
  C77} (2017) 82.

\bibitem{Aartsen:2017ulx}
{\bf IceCube} Collaboration, M.~G. Aartsen et~al., {\em Eur. Phys. J.} {\bf
  C77} (2017) 627.

\bibitem{Aartsen:2016pfc}
{\bf IceCube} Collaboration, M.~G. Aartsen et~al., {\em Eur. Phys. J.} {\bf
  C76} (2016) 531.

\bibitem{Aartsen:2018mxl}
{\bf IceCube} Collaboration, M.~G. Aartsen et~al., {\em Eur. Phys. J.} {\bf
  C78} (2018) 831.

\bibitem{Aartsen:2013dxa}
{\bf IceCube} Collaboration, M.~G. Aartsen et~al., {\em Phys. Rev.} {\bf D88}
  (2013) 122001.

\bibitem{antaresi3dmpos2019}
{\bf ANTARES, IceCube} Collaboration,  \pos{PoS(ICRC2019)522} (these
  proceedings).

\bibitem{hesedmpos2019}
{\bf IceCube} Collaboration,  \pos{PoS(ICRC2019)839} (these proceedings).

\bibitem{TheIceCube:2016oqi}
{\bf IceCube} Collaboration, M.~G. Aartsen et~al., {\em Phys. Rev. Lett.} {\bf
  117} (2016) 071801.

\bibitem{Aartsen:2017bap}
{\bf IceCube} Collaboration, M.~G. Aartsen et~al., {\em Phys. Rev.} {\bf D95}
  (2017) 112002.

\bibitem{Aartsen:2015exf}
{\bf IceCube} Collaboration, M.~G. Aartsen et~al., {\em Eur. Phys. J.} {\bf
  C76} (2016) 133.

\bibitem{Aartsen:2017ibm}
{\bf IceCube} Collaboration, M.~G. Aartsen et~al., {\em Nature Phys.} {\bf 14}
  (2018) 961--966.

\bibitem{Aartsen:2017xtt}
{\bf IceCube} Collaboration, M.~G. Aartsen et~al., {\em Phys. Rev.} {\bf D97}
  (2018) 072009.

\bibitem{Aartsen:2017nmd}
{\bf IceCube} Collaboration, M.~G. Aartsen et~al., {\em Phys. Rev. Lett.} {\bf
  120} (2018) 071801.

\bibitem{Li:2017dbe}
{\bf Super-Kamiokande} Collaboration, Z.~Li et~al., {\em Phys. Rev.} {\bf D98}
  (2018) 052006.

\bibitem{Agafonova:2018auq}
{\bf OPERA} Collaboration, N.~Agafonova et~al., {\em Phys. Rev. Lett.} {\bf
  120} (2018) 211801. [Erratum: Phys. Rev. Lett.121,no.13,139901(2018)].

\bibitem{Aartsen:2019tjl}
{\bf IceCube} Collaboration, M.~G. Aartsen et~al., {\em Phys. Rev.} {\bf D99}
  (2019) 032007.

\bibitem{Aartsen:2014njl}
{\bf IceCube} Collaboration, M.~G. Aartsen et~al.,
  \href{http://arxiv.org/abs/1412.5106}{{\tt arXiv:1412.5106}}.

\bibitem{upgradepos2019}
{\bf IceCube} Collaboration,  \pos{PoS(ICRC2019)1031} (these proceedings).

\bibitem{upgradedmpos2019}
{\bf IceCube} Collaboration,  \pos{PoS(ICRC2019)506} (these proceedings).

\end{thebibliography}\endgroup

%

\end{document}